\documentclass[11pt,onecolumn,twoside,draftcls]{IEEEtran}
\usepackage[centertags]{amsmath}
\usepackage{amsfonts}
\usepackage{amssymb}
\usepackage{graphicx,color}

\newcommand{\X}{{\mathbf X}}

\newcommand{\Y}{{\mathbf Y}}
\newcommand{\prob}{\text{Pr}}
\newcommand{\expec}{{\mathsf E}}
\newcommand{\peab}{P_{e|{\cal A},{\cal B}}} 
\newcommand{\qed}{\hspace*{\fill}~{\rule{2mm}{2mm}}\par\endtrivlist\unskip}
\newcommand{\tends}{\rightarrow}

\newcommand{\prcavg}{\bar{P}_{e,rc}}
\newcommand{\pe}{\bar{P}_{e}}
\newcommand{\al}{\mathsf{A}}

\newtheorem{lemma}{Lemma}
\newtheorem{theorem}{Theorem}

\begin{document}




%

\title{A Channel Coding Perspective of Collaborative Filtering \footnote{Preliminary results related to this submission were presented by us in \cite{us_isit09} (ISIT 2009, Seoul, Korea).}}
\author{
\authorblockN{S.\ T.\ Aditya \\}
\authorblockA{Department of Electrical Engineering\\
Indian Institute of Technology Bombay \\
Mumbai, India\\
{ Email:} staditya@ee.iitb.ac.in} \\
\and
\vspace*{0.8cm}
\authorblockN{Onkar Dabeer \\}
\authorblockA{School of Technology and Computer Science \\
Tata Institute of Fundamental Research\\
Mumbai, India \\
{ Email:} onkar@tcs.tifr.res.in} \\ 
\and
\vspace*{0.8cm}
\authorblockN{Bikash Kumar Dey\\}
\authorblockA{Department of Electrical Engineering \\
Indian Institute of Technology Bombay \\
Mumbai, India\\
{ Email:} bikash@ee.iitb.ac.in}
}
\maketitle

\begin{abstract}
We consider the problem of collaborative filtering from a channel coding perspective. We model the underlying rating matrix as a finite alphabet matrix with block constant structure. The observations are obtained from this underlying matrix through a discrete memoryless channel with a noisy part representing noisy user behavior and an erasure part representing missing data. Moreover, the clusters over which the underlying matrix is constant are {\it unknown}. We establish a sharp threshold result for this model: if the largest cluster size is smaller than $C_1 \log(mn)$ (where the rating matrix is of size $m \times n$), then the underlying matrix  cannot be recovered with any estimator, but if the smallest cluster size is larger than $C_2 \log(mn)$, then we show a polynomial time estimator with diminishing probability of error. In the case of uniform cluster size, not only the order of the threshold, but also the constant is identified.
\end{abstract}

\section{Introduction}

As new content mushrooms at a brisk pace, finding relevant information is increasingly a challenge. Consequently, recommendation systems are commonly being used to assist users: Amazon recommends books, Netflix recommends movies, LinkedIn recommends professional contacts, Google recommends webpages for a given query, etc. Such recommendation systems exploit various aspects to make suggestions: popularity amongst peers, similarity of content, available user-item ratings, etc. This paper is about collaborative filtering using the rating matrix: we are interested in making recommendations using only available ratings given by users to the items they have experienced. In a practical system, such a rating based collaborative filter is typically complemented by content-based analysis specific to the data.

There is vast literature on recommendation systems and collaborative filtering; see for example the special issue \cite{sp.issue} and the survey paper \cite{surveypaper}. Given the massive datasets and the lack of good statistical model of user behavior, the dominant stream of work has been to propose methods and demonstrate their scalability on real data sets. However, recently the Netflix Prize \cite{netflix} has popularized the problem to other research communities and several researchers have started exploring {\it provably} good methods. This paper falls in the latter category: we deal with fundamental limits of collaborative filters. In the remainder of this section, we first discuss related models and results, and then outline our model and results.

\subsection{Related Work}
The Netflix data consists of rating matrix where the rows correspond to movies and the columns correspond to users. Only a small fraction of the entries are known  and the goal is to estimate the missing entries, that is, this is a matrix completion problem. Several algorithms have been proposed and tested on this data set; see for example \cite{koren}. Mathematically, without any further restriction, this is an ill-posed problem. Motivated by this, some authors have recently considered the matrix completion problem under the restriction of low-rank matrices. (This problem also arises in other contexts such as location estimation in sensor networks.) This problem has attracted much attention, and in the past year a number of results have been reported. In \cite{recht2}, using nuclear norm minimization proposed in \cite{recht1}, an upper bound on the number of samples needed for recovery asymptotically is derived in terms of the size and rank of the matrix. In \cite{candes_tao}, a lower bound is established on the number of samples needed by any algorithm. The order of this lower bound is shown to be achievable in \cite{keshavan_isit09}. In \cite{bresler_admira}, the problem of matrix recovery from linear measurements (of which sampling is a special case) is considered and a new algorithm is proposed. In \cite{candes_plan}, the problem of matrix completion under bounded noise is considered. A semi-definite programming based algorithm is proposed and shown to have recovery error proportional to the noise magnitude. 

In this paper, we take an alternative channel coding viewpoint of the problem. Our results differ from the above works in several aspects outlined below.
\begin{itemize}
\item We consider finite alphabet for the ratings and a different model for the rating matrix based on row and column clusters.
\item We consider noisy user behavior, and our goal is not to complete the missing entries, but to estimate an underlying ``block constant" matrix (in the limit as the matrix size grows).
\item Since we consider a finite alphabet, even in the presence of noise, error free recovery is asymptotically feasible. Hence, unlike \cite{candes_plan}, which considers real-valued matrices, we do not allow any distortion. 
\end{itemize}
We next outline our model and results.

\subsection{Summary of Our Model and Results}

We consider a finite alphabet for the ratings. In this section, we briefly outline our model and results without any mathematical details; the details can be found in subsequent sections.

To motivate our model, consider an ideal situation where every user rates every item without any noise. In this ideal scenario, it is reasonable to expect that similar users rate  similar items by the same value.  We therefore assume that the users (items) are clustered into groups of similar users (items, respectively). The rating matrix in this ideal situation (say $\X$ with size $m \times n$) is then a block constant matrix (where the blocks correspond to cartesian product of row and column clusters). The observations are obtained from $\X$ by passing its entries through a discrete memoryless channel (DMC) consisting of an erasure channel modeling missing data and a noisy DMC representing noisy user behavior. Moreover, the row and column clusters are {\it unknown}.
The goal is to make recommendations by estimating $\X$ based on the observations. The performance metric we use is the probability of block error: we make an error if any of the entries in the estimate is erroneous. Our goal is to identify conditions under which error free recovery is possible in the limit as the matrix size grows large. Thus we view the recommendation system problem as a channel coding problem.

The cluster sizes in our model represent the resolution: the larger the cluster, the smaller are the degrees of freedom (or rate of the channel code). If the channel is more noisy and the erasures are high, then we can only support a small number of codewords. The challenge is to find the exact order. For our model, we show that if the largest cluster size (defined precisely in Section \ref{sect:bin}) is smaller than $C_1 \log(mn)$, where $C_1$ is a constant dependent on the channel parameters, then for any estimator the probability of error approaches one. On the other hand, if the smallest cluster size (defined precisely in Section \ref{sect:bin}) is larger than $C_2 \log(mn)$, where $C_2$ is a constant dependent on the channel parameters, then we give a polynomial time algorithm that has diminishing probability of error. Thus we identify the order of the threshold exactly. In the case of uniform cluster size, the constants $C_1$ and $C_2$ are identical and thus in this special case, even the constant is identified precisely. Moreover, for the special case of binary ratings and uniform cluster size, the algorithm used to show the achievability part does nor depend on the cluster size, erasure parameter, and needs knowledge of a worst case parameter for the noisy part of the channel. These results are obtained by averaging over $\X$ (as per the probability law specified in Section \ref{sect:model}).

\begin{figure}[t]
\begin{center}
\includegraphics[width=\textwidth]{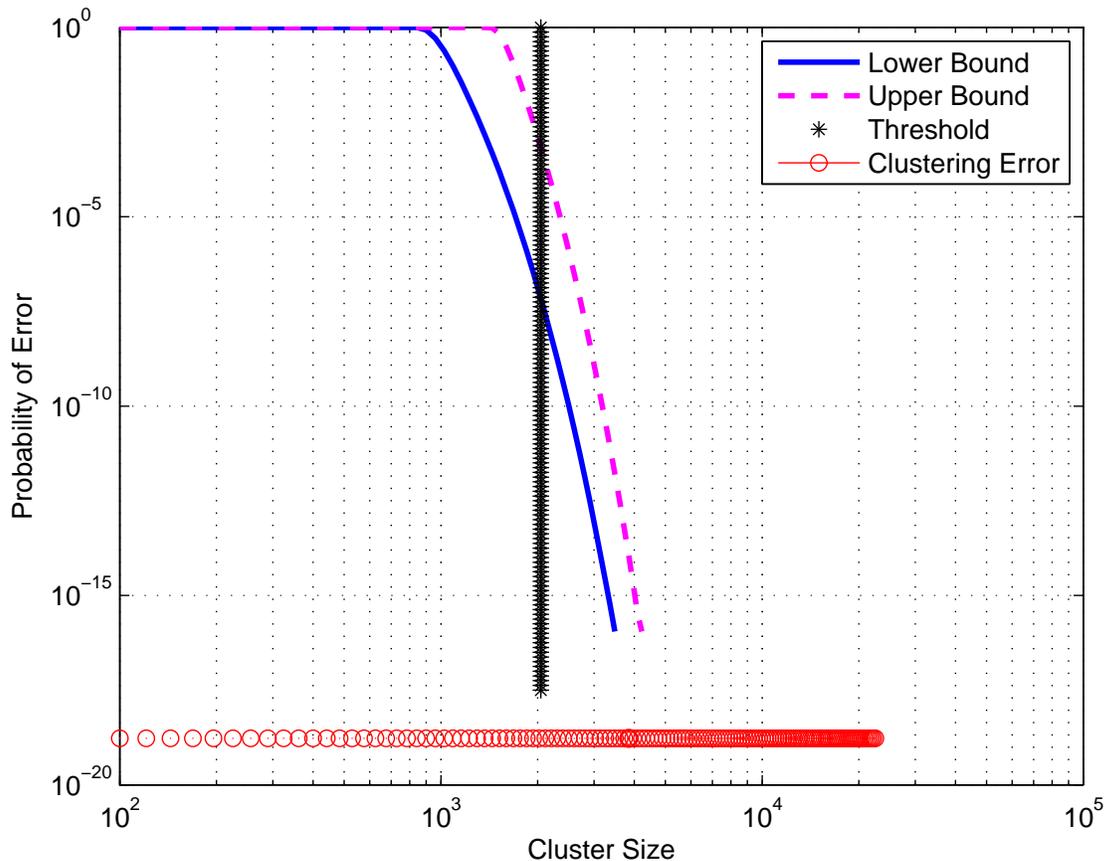}
\caption{The figure shows lower and upper bounds for the probability of error under known clustering (Theorem \ref{thm:bin_knownC}), the asymptotic cluster size threshold from Theorem \ref{thm:bin_main}, and an upper bound on the clustering error (Theorem \ref{thm:bin_C}) for the case $m=n=10^6$, erasure probability $\epsilon=0.9$, and binary symmetric channel with error $p=0.25$. The threshold in the clustering algorithm is chosen to be $d_0=(2p(1-p)+1)/3=0.4583$.}
\label{fig:bounds}
\end{center}
\end{figure}

The achievability part of our result is shown by first clustering the rows and columns, and then estimating the matrix entries assuming that the clustering is correct. The clustering is done by computing a normalized Hamming metric for every pair of rows and comparing with a threshold to determine if the rows are in same cluster or not. The converse is proved by considering the case when the clusters are known exactly. Our results for the average case show that the threshold is determined by the problem of estimating entries, and relatively, clustering is an easier task (see Figure \ref{fig:bounds} for an illustration). 

\subsection{Organization of the Paper}
The precise model for $\X$ and the observations is stated in Section \ref{sect:model}. The case of uniform cluster size and binary ratings leads to sharper bounds and results. Hence results for this case are given in Section \ref{sect:bin}. The case of general alphabets and non-uniform cluster sizes is considered in Section \ref{sect:general}. The conclusion is given in Section \ref{sect:con}, while all the proofs are collected together in Section \ref{sect:proofs}. 

\subsection{Notation}
All the logarithms are to the natural base unless specified otherwise. $D(\mu\|\nu)$ denotes the KL divergence (\cite{tmc}) between probability mass functions $\mu$ and $\nu$. By $T=\Omega(f(n))$ we mean that for $n$ large enough, $ T \geq \text{constant} \cdot f(n)$. By $1(A)$ we denote the indicator variable, which is 1 if $A$ is true and 0 otherwise.
 
\section{Model and Assumptions}
\label{sect:model}

The main elements of our model are a block constant ensemble of rating matrices (whose blocks of constancy are not known) and an observation matrix obtained from the underlying rating matrix via a noisy channel and erasures. The noise in the observations represents the inherent noise in user-item ratings as well as the error in our model. The erasures denote missing entries. To be more precise,  suppose $\X$ is the unknown $m \times n$ rating matrix with entries from a finite alphabet, where $n$ is the number of buyers and $m$ is the number of items. Let ${\cal A}=\{A_i\}_{i=1}^r$ and ${\cal B}=\{B_j\}_{j=1}^t$ be partitions of $[1:m]$ and $[1:n]$ respectively.  We call the sets $A_i \times B_j$ clusters and we call $A_i$'s ($B_j$'s) the row (column) clusters. We denote the corresponding row and column cluster sizes by $m_i$ and $n_j$, and the number of row clusters and the number of column clusters by $r$ and $t$ respectively. Thus $\sum_{i=1}^r m_i = m$, $\sum_{j=1}^t n_j = n$.

 We state our results under two sets of conditions - the set of conditions A1)-A4) and B1)-B3) below. Conditions A1)-A4) are a special case of conditions B1)-B3). The results under A1)-A4) are sharper and illustrate the important concepts more easily. Hence they are stated separately.  We begin by stating and discussing A1)-A4) first and then we state B1)-B3). (A few additional conditions needed in the results are stated at appropriate places.)\\

\noindent
{\bf Conditions A1)-A4):}
The conditions A1)-A4) below correspond to binary rating matrix with equal size clusters and uniform probability of sampling entries.
\begin{enumerate}
\item[A1)] The entries of $\X$ are from $\{0,1\}$. 
\item[A2)] The row (column) clusters are of equal size: $m_i=m_0$, $n_i = n_0$ for all $i$.
\item[A3)] $\X$ is constant over the cluster $A_i \times B_j$ and the entries are  i.i.d.\ Bernoulli(1/2) across the clusters.
\item[A4)] The observed data $\Y \in \{0,1,e\}$ ($e$ denotes erasure) is obtained by passing the entries of $\X$ through the cascade of a binary symmetric channel (BSC) with probability of error $p$ and an erasure channel with erasure probability $\epsilon$.
\end{enumerate}
The cluster sizes are representative of the {\it resolution} of $\X$ - large cluster sizes correspond to a coarse structure with fewer degrees of freedom in choosing $\X$, while small cluster size corresponds to a fine structure.
Condition A2) suggests that we can think of the cluster size $m_0 n_0$ as representative of the resolution of $\X$ and it plays a central role in our results. If we think of all permissible $\X$ as a channel code, then a higher $m_0 n_0$ corresponds to a smaller rate code. 
However, in order to interpret $m_0 n_0$ precisely, we also need to take into account condition A3). When the entries of the cluster are filled with i.i.d.\ Bernoulli(1/2) random variables as per A3), it is likely that rows in two clusters turn out to be the same, and hence these two row clusters can be merged to form a single bigger cluster.  The following lemma shows that if the number of clusters is  $\Omega(\log(n))$, then this happens with small probability and hence we should think of $m_0 n_0$ as {\it the} representative cluster size.
\begin{lemma}
\label{lem:mn}
If $t \geq (2+\delta) \log_2(n)$, $\delta > 0$, then 
\[
P\left(\text{Rows in two different clusters are same}\right) \leq \frac{1}{m^\delta}
\] 
and a similar result holds for the column clusters.
\end{lemma}
\proof
Each row is uniformly distributed over $2^t$ possibilities and rows in different clusters are independent. Hence the probability that any given pair of rows is same is $1/2^t$. Since there are $\binom{r}{2}$ pairs, we then have 
\[
P\left(\text{Rows in two different clusters are same}\right)\leq \frac{\binom{r}{2}}{2^t},
\]
Since $r\leq m$, we have
\[
P\left(\text{Rows in two different clusters are same}\right)\leq \frac{m^2}{2^t}.
\] 
Hence if $t>(2+\delta)\log_2m$ for some $\delta>0$, then
\[
P\left(\text{Rows in two different clusters are same}\right)\leq \frac{1}{m^\delta}.
\]
\qed

Condition A3) also implies that in any row or column, for large matrices, roughly the number of 0s and 1s is same. This essentially implies that the opinions are diverse for any user or item. While this may seem unrealistic (and can indeed be fixed), we prefer the Bernoulli(1/2) model for the following reason: under this assumption no recommendations can be extracted from any row or column alone and thus collaborative filtering is necessary. Such a model is desirable for evaluation of collaborative filtering schemes. Moreover, one can pre-process data so that rows and columns with fraction of 1s far from 1/2 are removed (because they are relatively easy to recommend) and then assumption A3) is reasonable. We note that in condition A3), we only specify the probability law of $\X$ given the clusters; the clusters are deterministic, even though they are unknown.

The BSC in A4) models the inherent noise in user-item ratings as well as modeling error, while the erasure channel models the missing data.  \\

\noindent
{\bf Conditions B1)-B3):} These conditions are more general allowing any finite alphabet and non-uniform cluster sizes.
\begin{enumerate}
\item[B1)] The entries of $\X$ are from a finite alphabet $\al$. 
\item[B2)] $\X$ is constant over the cluster $A_i \times B_j$ and the entries across the clusters are  i.i.d.\ with a uniform distribution over $\al$.
\item[B3)] The observed data $\Y \in \al\cup\{e\}$ ($e$ denotes erasure) is obtained from $\X$ as follows 
\begin{enumerate}
\item The entries of $\X$ are passed through a DMC with probability law $q(.|.)$ and output alphabet $\al$, resulting in $\tilde{\X}$. 
\item  The entries $\tilde{\X}_{ij}$ are then passed through an erasure channel with erasure probability $\epsilon$.
\end{enumerate}
\end{enumerate}

\section{Binary Rating Matrix}
\label{sect:bin}

In this section, we state our results under conditions A1)-A4). The main result of this section appears in Section \ref{sect:bin_main}. It is obtained by studying two quantities: probability of error when the clustering is {\it known} (Section \ref{sect:bin_knownC}) and probability error in clustering for a specific algorithm (Section \ref{sect:bin_C}). 

\subsection{Main Result}
\label{sect:bin_main}

Our main result stated below identifies a threshold on the cluster size above which error free recovery is asymptotically feasible but below which error free recovery is not possible. 
\begin{theorem}
\label{thm:bin_main}
Suppose conditions A1)-A4) are true and the clusters are {\it unknown}. Let $p_1 = \epsilon+2(1-\epsilon)\sqrt{p(1-p)}$.  
Suppose that $\epsilon < 1$ and $ p \in [0,p_0]$, $p_0 < 1/2$.
\begin{enumerate}
\item {\bf Converse:} If 
\[
m_0 n_0 < (1-\delta) \frac{\ln(mn)}{\ln(1/p_1)}, \quad \delta > 0,
\]
then $\pe \tends 1$ for {\it any}  estimator.
\item {\bf Achievability:} If $t =\Omega(\log(n))$, $r =\Omega(\log(m))$, $\limsup m/n < \infty$, $\limsup n/m < \infty$ and
\[
m_0 n_0 > \frac{\ln(mn)}{\ln(1/p_1)},
\] 
then $\pe \tends 0$ for the following polynomial time estimator:
\begin{itemize}
\item Cluster rows and columns using the algorithm of Section \ref{sect:bin_C} using the threshold $d_0 \in (2p_0 (1-p_0), 1/2)$ (which does not depend on $\epsilon, m_0, n_0$).   
\item Employ majority decoding in a cluster (as in Section \ref{sect:bin_knownC}) assuming the clustering to be correct.
\end{itemize}
\end{enumerate}
\end{theorem}
\proof The proof is given in Section \ref{sect:proof_bin_main}. \qed

The result identifies $\ln(mn)/\ln(1/p_1)$ as the cluster size threshold.
The first part states that if the cluster size is too small, then any estimator makes an error with high probability.  The second part states that if the cluster size is large enough, then diminishing probability of error can be achieved with a polynomial time estimator, which does not need knowledge of $\epsilon, m_0, n_0$ and needs only knowledge of a worst case bound on $p$. The result is reminiscent of the channel coding theorem in the context of our model.

The proof of Part 1) of Theorem \ref{thm:bin_main} relies on lower bounding $\pe$ by considering the case of known clustering (see Theorem \ref{thm:bin_knownC} in Section \ref{sect:bin_knownC}). The proof of Part 2) of Theorem \ref{thm:bin_main} relies on showing that for the average case, the probability of error in clustering is much smaller than the probability of error in filling values when the clusters are known (see Theorem \ref{thm:bin_C} in Section \ref{sect:bin_C}). We illustrate this in Figure \ref{fig:bounds} by plotting various bounds: for $m=n=10^6$, $m_0=n_0$ ranging from 10 to 150, $p=0.25$ and $\epsilon=0.9$, we plot 
\begin{itemize}
\item upper and lower bounds for probability of error when clustering is known (from Theorem \ref{thm:bin_knownC}),  
\item upper bound on probability of clustering error (from Theorem \ref{thm:bin_C}),
\item and the asymptotic threshold $\ln(mn)/\ln(1/p_1)$ (from Theorem \ref{thm:bin_main}).
\end{itemize}
It is seen that around the asymptotic threshold, the probability of clustering error is dominated by the probability of error in filling values under known clustering.

\subsection{Known Clustering}
\label{sect:bin_knownC}

In this section, we consider the case when the clusters are known. Under this assumption, the decoder only has to estimate the value in a cluster, and the minimum probability of error estimator under A3) is just a majority decoder. The analysis of this decoder is elementary and we state a stronger result for a fixed $\X$ with possibly unequal cluster sizes. Let
\[
s_{\ast}(\X) := \min_{i,j} m_i(\X) n_j(\X), \quad s^{\ast}(\X) := \max_{i,j} m_i(\X) n_j(\X),
\]
where $\{m_i(\X)\}$ and $\{n_j(\X)\}$ are the row and column cluster sizes in $\X$.
\begin{theorem}
\label{thm:bin_knownC}
Suppose conditions A1), A3) are true and in addition assume that the clusters are known. Let
$$s_{\ast}(\X) \geq \frac{\ln(2)}{\ln(1/p_1)}.$$
Then the probability of error in filling in values satisfies
\begin{equation}
\label{eq:lb-ub3}
\begin{split}
\peab(\X) & \geq 1-\exp\left(- \frac{1}{4} \sqrt{\frac{p}{1-p}}\frac{m n p_1^{s^{\ast}(\X)}}{s^{\ast}(\X)(s^{\ast}(\X)+1)}\right), \\
\peab(\X) & \leq 1-\exp\left(- \frac{2\ln(2)m n p_1^{s_{\ast}(\X)}}{s_{\ast}(\X)}\right).
\end{split}
\end{equation}
Suppose we are given a sequence of rating matrices of increasing size, that is, $mn \tends \infty$. Then the following are true.
\begin{enumerate}
\item If 
\[
s_{\ast}(\X) \geq \frac{\ln(mn)}{\ln(1/p_1)} 
\]
then $\peab(\X) \tends 0$.
\item If 
\[
 s^{\ast}(\X) \leq \frac{(1-\delta)\ln(mn)}{\ln(1/p_1)} \text{ for some } \delta > 0,
\]
then $\peab(\X) \tends 1$.
\end{enumerate}
\end{theorem}
\proof The proof is given in Section \ref{sect:proof_bin_knownC}. \qed

We note that when all the clusters are of the same size (which happens with high probability as per Lemma \ref{lem:mn}), then the above result states that there is a sharp threshold: if the cluster size is smaller than $\ln(mn)/\ln(1/p_1)$, then exact recovery is not possible, but if it is larger, then we can make probability of error as small as we wish. \\

\noindent
{\bf Example:}
For $m=n=10^6$, $m_i=n_j=n_0$, $\epsilon=0.9$, $p=0.25$, this threshold corresponds to clusters of size about $45  \times 45 = 2025$. We plot the lower and upper bounds for $\peab(\X)$ from Theorem \ref{thm:bin_knownC}  and the threshold in Figure \ref{fig:bounds}. \\

\noindent
{\bf Remark:} A finer analysis reveals that we can refine Part 2) of Theorem \ref{thm:bin_knownC} (and hence also Part 1) of Theorem \ref{thm:bin_main}) by letting $\delta$ approach zero as $m,n\tends \infty$. The result holds as long as $\delta_{m,n} \ln(mn) - 2\ln\ln(mn) \tends \infty$.

\subsection{Probability of Clustering Error}
\label{sect:bin_C}

To get an upper bound on the probability of error $\pe$, in this section we analyze a specific collaborative filter: we first cluster the rows and columns using the algorithm described below and then we fill in values using the majority decoder assuming that the clustering is correct. The majority decoder has already been analyzed in Section \ref{sect:bin_knownC} and for proving Part 2 of Theorem \ref{thm:bin_main}, we only need to analyze the probability of error in clustering. 

\noindent
{\bf Clustering Algorithm:} We cluster rows and columns separately. 
For  rows $i,j$, the normalized Hamming distance over commonly sampled entries is 
\[
d_{ij} = \frac{1}{N_{ij}} \sum_{k=1}^n 1\left(Y_{ik} \neq e, Y_{jk} \neq e, Y_{ik} \neq Y_{jk}\right),
\]
where $N_{ij}$ is the number of commonly sampled positions in rows $i$ and $j$, given by
\[
N_{ij} = \sum_{k=1}^n 1\left(Y_{ik} \neq e, Y_{jk} \neq e\right).
\]
Let $I_{ij}$ be equal to 1 if rows $i$, $j$ belong to the same cluster and let it be 0 otherwise. The algorithm gives an estimate:
\begin{equation*}
\hat{I}_{ij} = 
\begin{cases}
 & 1, \quad d_{ij} < d_0, \\
& 0, \quad d_{ij} \geq d_0,
\end{cases}
\end{equation*}
where $d_0$ is a treshold whose choice will be discussed later. A similar algorithm is used to cluster columns.
We are interested in the probability that we make an error in row clustering averaged over the probability law on the rating matrices defined as
\[
\prcavg = \prob\left(\hat{I}_{ij} \neq I_{ij} \text{ for some } i,j\right).
\]
We note that this is a conservative definition of clustering error. As seen in Lemma \ref{lem:mn}, there is a small chance that rows in different clusters may be the same resulting in the merging of two clusters into a larger one. The above definition of error does not account for this and declares more errors. We use this conservative definition of clustering error to simplify analysis.

\begin{theorem}
\label{thm:bin_C}
Suppose conditions A1)-A4) are true. Let $r_1>1$, $r_2 \in (0,1)$ be constants and let $h_\ast$ be the smaller root of the quadratic equation
\begin{equation}
\label{eq:opt_h}
	2\mu\nu(1-d_0)h^2+(2d_0-2\mu\nu-1)h+1-2d_0=0, 
\end{equation}
where $\mu:= 2p(1-p), \nu=1-\mu$.
Suppose the threshold $d_0 \in (\mu, \mu+1/2)$. Let
\begin{align*}
& \alpha_1  = D\left(r_1(1-\epsilon)^2||(1-\epsilon)^2\right), \\
& \alpha_2  = D\left(r_2(1-\epsilon)^2||(1-\epsilon)^2\right),  \\
& \lambda_1(n_0)  =\frac{1}{2}\left(1+\left(\frac{1-\nu h^\ast}{1-\mu h^\ast}\right)^{n_0r_2(1-\epsilon)^2}\right), \text{ and,}\\
& \lambda_2(n_0) =\frac{1}{2}\left(1+2^{-n_0\alpha_2}\right).
\end{align*}
Then for the above clustering algorithm,
\[
\prcavg \leq \frac{m(m-1)}{2} \max\left\{P(\hat{I}_{ij}=0\big| I_{ij}=1), P(\hat{I}_{ij}=1\big| I_{ij}=0)\right\},
\]
where
\begin{equation}
P(\hat{I}_{ij}=0\big| I_{ij}=1)  \leq \exp\left(- n \min\left\{r_2(1-\epsilon)^2 D(d_0 \| \mu), \alpha_2\right\} \right)
\end{equation}
and
\begin{align}
& P(\hat{I}_{ij}=1\big| I_{ij}=0)  \leq \min\left\{ P_1, P_2\right\} \\
& P_1  = \left(\frac{1-\mu h_\ast}{(1-h_\ast)^{d_0}}\right)^{nr_1(1-\epsilon)^2} \lambda_1(n_0)^t + \exp(-\alpha_1 n) + \lambda_2(n_0)^t \\
& P_2  = \exp\left(-\alpha_3t\right)
\end{align}
for a positive constant $\alpha_3$.
\end{theorem} 
\proof The proof is given in Section \ref{sect:proof_bin_C}. \qed

The proof uses the union bound and considers pairwise errors. The pairwise errors consists of two cases: error when the pair of rows is in the same cluster and error when they are in different clusters. The probability of the first kind of error is exponentially decaying in $n$. The probability of the second kind of error is upper bounded by the minimum of $P_1$ and $P_2$: while $P_1$ is tight for finite $n$ and large $p,\epsilon$, the bound $P_2$ is useful for establishing asymptotic results (like Theorem \ref{thm:bin_main}) for {\it all} $p,\epsilon$. For example, in Figure \ref{fig:bounds}, the upper bound on clustering error is dominated by $P_1$, while the proof of Part 2) of Theorem \ref{thm:bin_main} uses $P_2$. We note that both $P_1$ and $P_2$ have terms that decay exponentially in $n$ as well as $t$. The terms decaying exponentially in $t$ are related to Lemma \ref{lem:mn} and the conservative definition of clustering error as discussed before the statement of Theorem \ref{thm:bin_C}. These terms are the origin of the $t=\Omega(\log(n))$ condition in Part 2) of Theorem \ref{thm:bin_main} and can perhaps be avoided with more sophisticated analysis; however, we prefer to work with this condition since as per Lemma \ref{lem:mn}, the 
condition $t=\Omega(\log(n))$ is anyway needed for interpreting $m_0n_0$ as the representative cluster size.

\section{General Finite Alphabet and Non-uniform Clusters}
\label{sect:general}

In this section, we consider a general finite alphabet $\al$ and non-uniform cluster sizes. We work with assumptions B1)-B3) described in the Section \ref{sect:model} and generalize the results in Section \ref{sect:bin}. To state our results, we first introduce some notation. For $p,q \in \al$, define
\begin{equation}
	\label{eq:mu}
	\mu_{pq} :=\sum_{y_i \neq y_j}q(y_i|p)q(y_j|q).
\end{equation}
If $A_1$, $A_2$ are i.i.d.\ uniform on $\al$ and we pass them through the DMC $q(\cdot |\cdot)$ to get outputs $\tilde{A}_1$, $\tilde{A}_2$, then
\begin{align*}
\prob(\tilde{A}_1 \neq \tilde{A}_2 | A_1 = A_2) & = \frac{1}{|\al|}\sum_{p \in \al} \mu_{pp} =: d_{lb} \\
\prob(\tilde{A}_1 \neq \tilde{A}_2) & = \frac{1}{|\al|^2}\sum_{p,q \in \al} \mu_{pq} =: d_{ub}.
\end{align*}
The following useful lemma sheds light on the relationship between $d_{lb}$ and $d_{ub}$.
\begin{lemma}
\label{lem:dineq}
For any DMC, $d_{ub} \geq d_{lb}$,
with equality iff $q(y|p)=q(y|q)$ $\forall \ \ p, q, y \in \al$. 
\end{lemma}
\proof The proof is given in Section \ref{proof:lem:dineq}. \qed

We next state our main result for general finite alphabet and non-uniform cluster size.

\begin{theorem}
\label{thm:dmc_main}
Suppose conditions B1)-B3) are true and the clusters are {\it unknown}. Then there exist constants $p_1, p_2 \in (0,1)$, $p_1 > p_2$ such that 
\begin{enumerate}
\item {\bf Converse:} If 
\[
\max_{i,j} m_i n_j < (1-\eta) \frac{\ln(mn)}{\ln(1/p_2)}, \quad \eta > 0,
\]
then $\pe \tends 1$ for {\it any}  estimator.
\item {\bf Achievability:} Suppose that there exist some $y, p, q \in \al$ such that $p\neq q$ and $q(y|p) \neq q(y|q)$. (By Lemma \ref{lem:dineq}, this ensures that $d_{lb} < d_{ub}$.) If $n^2/(n_1^2+n_2^2+\ldots+n_t^2) =\Omega(\log(m))$, $m^2/(m_1^2+m_2^2+\ldots+m_r^2) =\Omega(\log(n))$, $\limsup m/n < \infty$, $\limsup n/m < \infty$  and
\[
\min_{i,j} m_i n_j > \frac{\ln(mn)}{\ln(1/p_1)},
\] 
then $\pe \tends 0$ for the following polynomial time estimator:
\begin{itemize}
\item Cluster rows and columns using the algorithm of Section \ref{sect:bin_C} using the threshold $d_0 \in (d_{lb}, d_{ub})$ (which does not depend on $\epsilon, m_i, n_j$).   
\item Employ maximum likelihood decoding in a cluster assuming the clustering is correct.
\end{itemize}
\end{enumerate}
\end{theorem}
\proof The proof is similar to Theorem \ref{thm:bin_main}; we now use Theorems \ref{thm:dmc_knownC} and \ref{thm:gen_clus} in place of Theorems \ref{thm:bin_knownC} and \ref{thm:bin_C} respectively. \qed

The above result again identifies $\ln(mn)$ as the exact order of the cluster size threshold for asymptotic recovery. Similar to the binary alphabet and uniform cluster size case in Section \ref{sect:bin}, the constants $p_1$, $p_2$ arise from the case when the clusters are known (see Theorem \ref{thm:dmc_knownC} below). The gap between the constants $p_1, p_2$ can be made arbitrarily small: the proof of Theorem \ref{thm:dmc_knownC} identifies a constant $C_1$ (see equation \eqref{eq:C_1}) such that for any $\delta>0$, 
\[
p_1 = \epsilon + (1-\epsilon)\exp(-C_1+\delta), \quad p_2 = \epsilon + (1-\epsilon)\exp(-C_1-\delta)
\]
is a valid choice in Theorem \ref{thm:dmc_main}. 

We next consider the case when the clusters are known and extend Theorem \ref{thm:bin_knownC}.
\begin{theorem}
\label{thm:dmc_knownC}
Suppose conditions B1)-B3) are true and in addition assume that the clusters are {\it known}. 
Also let 
\begin{equation}\label{eq:s-ast-x-lb}
	s_\ast(\X) \geq \frac{\ln(1/2|\al|)}{\ln(\epsilon/p_2)},
\end{equation}
where $p_1,p_2$ are as defined above.
Then for a sequence of rating matrices of increasing size $mn \tends \infty$, the following are true.
\begin{enumerate}
\item If 
\[
s_{\ast}(\X) \geq \frac{\ln(mn)}{\ln(1/p_1)}, 
\]
then $\peab(\X) \tends 0$.
\item If 
\[
 s^{\ast}(\X) \leq \frac{(1-\zeta)\ln(mn)}{\ln(1/p_2)}, \text{ for some } \zeta > 0,
\]
then $\peab(\X) \tends 1$.
\end{enumerate}
\end{theorem}
\proof The proof is given in Section \ref{sect:proof:thm:dmc_knownC}. \qed

Finally, we study the performance of the clustering algorithm and extend Theorem \ref{thm:bin_C}.
\begin{theorem}
\label{thm:gen_clus} Suppose conditions B1)-B3) are true and in addition suppose that there exist some $y, p, q \in \al$ such that $p\neq q$ and $q(y|p) \neq q(y|q)$. (By Lemma \ref{lem:dineq}, this ensures that $d_{lb} < d_{ub}$.)
 If we choose the threshold $d_0 \in (d_{lb}, d_{ub})$, then 
\begin{equation}
\label{eq:prc_gen}
\prcavg \leq c^{\prime} \frac{m(m-1)}{2} \exp\left(-cn^2/(n_1^2+n_2^2+\ldots+n_t^2)\right),
\end{equation}
for some positive constants $c$, $c^{\prime}$. Consequently, if $n^2/(n_1^2+n_2^2+\ldots+n_t^2) =\Omega(\log(m))$, then $\prcavg \tends 0$ as $m,n \tends \infty$.
\end{theorem}
\proof The proof is given in Section \ref{sect:proof:thm:gen_clus}. \qed

\section{Conclusion}
\label{sect:con}
We take a channel coding perspective of collaborative filtering and identify the threshold on cluster size for perfect reconstruction of the underlying rating matrix. The result is similar in flavor to some recent results in completion of real-valued matrices. The advantage of our model is that the proofs are relatively simple relying on Chernoff bounds and noisy user behavior can be easily handled. 

In the typical applications of recommendation systems, there is a lack of good models. We believe that our model has two characteristics that make it suitable for analytical comparison of various methods: a) in our model the user opinions are diverse and no single user/item reveals much information about itself, that is, collaborative filtering is necessary; b) as we have shown, the model is analytically tractable. There are several directions where this model may turn out to be useful: analysis of bit error probability instead of block error probability, analysis of local popularity based mechanisms, etc.

\section{Proofs of Results}
\label{sect:proofs}

\subsection{Proof of Theorem \ref{thm:bin_main}}
\label{sect:proof_bin_main}
The proof is based on Theorems \ref{thm:bin_knownC} and \ref{thm:bin_C}.

When ${\cal A}, {\cal B}$ are known, under our model all feasible rating matrices are equally likely. Hence the ML decoder gives the minimum probability of error and so we have
$
\pe \geq E[\peab(\X)].
$
To prove Part 1), we lower bound $E[\peab(\X)]$.
Let $T$ be the event that $s^{\ast}(\X)>m_0n_0$. Proceeding as in Lemma \ref{lem:mn}, we have for $t \geq (2+\delta) \log_2(n)$, $r \geq (2+\delta) \log_2(m)$, $\delta > 0$,
\[
\prob (T) \leq \frac{1}{m^\delta} + \frac{1}{n^\delta}.
\]
Hence $\prob (T) \tends 0$. Now,
\[
E[\peab(\X)]  \geq  E[\peab(\X); T^c]. 
\]
But on the event $T^c$, $s^{\ast}(\X)=m_0 n_0$ and hence we get
\begin{equation}
\pe \geq E[\peab(\X)] \geq (1-\prob (T))\peab(\X).
\end{equation}
But from Part 1) of Theorem \ref{thm:bin_knownC}, $\peab(\X) \tends 1$ for 
\[
m_0 n_0 < (1-\delta) \frac{\ln(mn)}{\ln(1/p_1)}, \quad \delta > 0.
\]
This proves Part 1).
 
Next we prove Part 2). Let $D$ denote the event that the clustering is identified correctly. We note that the probability of error in estimating $\X$ averaged over the probability law on the block constant matrices satisfies 
\begin{equation*}
\begin{split}
\pe & \leq E\left[\peab(\X) \prob(D)+ \prob(D^c)\right] \\
& \leq E\left[\peab(\X)\right]  + \left(\prcavg + \bar{P}_{e,cc}\right)
\end{split}
\end{equation*}
where $\bar{P}_{e,cc}$ is the probability of error in column clustering. The desired result follows from Part 2) of Theorem \ref{thm:bin_knownC} and Theorem \ref{thm:bin_C}. \qed

\subsection{Proof of Theorem \ref{thm:bin_knownC}}
\label{sect:proof_bin_knownC}
Suppose in cluster  $A_i \times B_j$ we have $s$ non erased samples. Then the probability of correct decision in this cluster is given by
\begin{equation}
\label{eq:eijs}
\begin{split}
\prob(E_{i,j,s}^c) & = \sum_{q=0}^{\lfloor \frac{s}{2}\rfloor}\binom{s}{q} p^q(1-p)^{s-q}  \text{ if }s\text{ is odd}\\
& = \sum_{q=0}^{\frac{s}{2}-1}\binom{s}{q}p^q(1-p)^{s-q} \\
& \quad +
\dfrac{1}{2}\binom{s}{\frac{s}{2}}p^{\frac{s}{2}}(1-p)^{\frac{s}{2}} \text{ if }s\text{ is even}.
\end{split}
\end{equation}
Averaging over the number of non erased samples, the probability of correct decision in cluster $A_i \times B_j$ is given by
\begin{equation}
\label{eq:eij}
\prob(E_{i,j}^c) =  \sum_{s=0}^{m_i n_j}\binom{m_i n_j}{s}\epsilon^{m_i n_j-s}(1-\epsilon)^s \prob(E_{i,j,s}^c). 
\end{equation}
Since the erasure and BSC are memoryless
\begin{align}
\peab(\X) & = \prob\left(\cup_{i=1,j=1}^{r,t} E_{i,j} \right) \nonumber \\
& = 1 - \prod_{i=1,j=1}^{r,t} \prob\left(E_{i,j}^c\right). \label{eq:error}
\end{align}
Equations \eqref{eq:eijs}, \eqref{eq:eij}, and \eqref{eq:error} specify the probability of error. 

\noindent {\bf Upper Bound:}
The desired upper bound is obtained by deriving a lower bound on $\prob(E_{i,j,s}^c)$. 
First we note that from \eqref{eq:eijs},
\[
1-\prob(E_{i,j,s}^c)  \leq \sum_{\lceil \frac{s}{2} \rceil}^s \binom{s}{q}p^q(1-p)^{s-q}. 
\]
But for $0\leq p \leq \frac{1}{2}$ and $q\geq \frac{s}{2}$, $p^q(1-p)^{s-q}\leq p^{\frac{s}{2}}(1-p)^{\frac{s}{2}}$. 
Substituting this in the previous equation, we have
\begin{equation}
	\prob(E_{i,j,s}^c) \geq 1-(2\sqrt{p(1-p)})^s. \label{eq:eijslb}
\end{equation}
From Equations \eqref{eq:eij} and \eqref{eq:eijslb}, we have $\prob(E_{i,j}^c) \geq 1- p_1^{m_in_j}$ and so from \eqref{eq:error}, 
\begin{equation*}
	\peab(\X)\leq 1- \prod_{i=1,j=1}^{r,t} \left(1-p_1^{m_i n_j}\right).
\end{equation*}
We note that for $x \in [0,1/2]$, $1-x \geq \exp(-2\ln(2) x)$. Hence
\begin{equation*}
\exp\left(-2\ln(2) \sum_{i=1,j=1}^{r,t} p_1^{m_i n_j} \right)  \leq \prod_{i=1,j=1}^{r,t} \left(1- p_1^{m_i n_j} \right).
\end{equation*}
Where the first inequality holds for $p_1^{m_i n_j} \leq 1/2$. This is true since $s_{\ast}(\X) \geq {\ln(2)}/{\ln(1/p_1)}$. The upper bound follows by noting that 
\[
\sum_{i=1,j=1}^{r,t} p_1^{m_i n_j} \leq r t p_1^{s_{\ast}(\X)} \leq \frac{mn}{s_{\ast}(\X)} p_1^{s_{\ast}(\X)}.
\]

\noindent
{\bf Lower Bound:}
The lower bound on $\peab(\X)$ is obtained from an upper bound on $\prob(E_{i,j,s}^c)$. From \eqref{eq:eijs}, 
\begin{align*}
1-\prob(E_{i,j,s}^c) &\geq \frac{1}{2}\binom{s}{\lceil s/2 \rceil}p^{\lceil s/2 \rceil}(1-p)^{s-\lceil s/2 \rceil},\\
&\geq \frac{1}{2(s+1)}2^{sh\left(\lceil s/2 \rceil /s\right)}p^{\lceil s/2 \rceil}(1-p)^{s-\lceil s/2 \rceil}.
\end{align*} 
If $s$ is even, we have
\begin{equation}
	1-\prob(E_{i,j,s}^c) \geq \frac{1}{2(s+1)}\left(2\sqrt{p(1-p)}\right)^s. \label{eijs-lb1}
\end{equation}
For $s$ odd, 
\[
h\left(\lceil s/2 \rceil /s\right)=h(1/2 + 1/2s) \geq 1-1/s^2,
\]
and so 
\begin{equation}
	1-\prob(E_{i,j,s}^c) \geq \frac{2^{-1/s}}{2(s+1)}\sqrt{\frac{p}{1-p}}\left(2\sqrt{p(1-p)}\right)^s. \label{eijs-lb2}
\end{equation}
From \eqref{eijs-lb1} and \eqref{eijs-lb2}, we have for all $s$,
\begin{equation}
	1-\prob(E_{i,j,s}^c) \geq \frac{1}{4(s+1)}\sqrt{\frac{p}{1-p}}\left(2\sqrt{p(1-p)}\right)^s. \label{eijs-lb}
\end{equation}
Now from \eqref{eq:eij},
\begin{align*}
\prob(E_{i,j}^c) &\leq 1- \frac{1}{4}\sqrt{\frac{p}{1-p}}\sum_{s=0}^{m_i n_j}\binom{m_i n_j}{s}\epsilon^{m_i n_j-s}\left((1-\epsilon)2\sqrt{p(1-p)}\right)^s \frac{1}{s+1} \\
&\leq 1- \frac{p_1^{m_in_j}}{4(m_in_j+1)}\sqrt{\frac{p}{1-p}}.
\end{align*}
Using this bound on $\prob(E_{i,j}^c)$ in \eqref{eq:error}, we have 
\begin{align}
\peab(\X) &\geq 1 - \prod_{i=1,j=1}^{r,t} \left(1- \frac{p_1^{m_in_j}}{4(m_in_j+1)}\sqrt{\frac{p}{1-p}}\right) \nonumber \\
&\geq 1- \exp\left(-\frac{1}{4}\sqrt{\frac{p}{1-p}}\sum_{i=1,j=1}^{r,t}\frac{p_1^{m_in_j}}{m_in_j+1}\right)   \label{eq:peab1} \\
&\geq 1- \exp\left(-\frac{1}{4}\sqrt{\frac{p}{1-p}}rt\frac{p_1^{s^{\ast}(\X)}}{s^\ast(\X)+1}\right) \nonumber \\
&\geq  1- \exp\left(-\frac{1}{4}\sqrt{\frac{p}{1-p}}mn\frac{p_1^{s^{\ast}(\X)}}{s^\ast(\X)+1}\right), \nonumber 
\end{align}
where in \eqref{eq:peab1} we have used $1-x \leq \exp(-x)$. This completes the proof of \eqref{eq:lb-ub3}.

\noindent
{\bf Asymptotics:}
Now consider a sequence of rating matrices of increasing size.
The upper bound on error in \eqref{eq:lb-ub3} is a decreasing function of $s_{\ast}(\X)$. 
Hence if  
\[
s_{\ast}(\X) \geq \frac{\ln(mn)}{\ln(1/p_1)}, 
\]then 
\[
\peab(\X) \leq \frac{2\ln (2) \ln(1/p_1)}{\ln(mn)} \tends 0.
\]
Now suppose
\[
 s^{\ast}(\X) \leq \frac{(1-\delta)\ln(mn)}{\ln(1/p_1)}, \text{ for some } \delta > 0.
\]
The lower bound on error \eqref{eq:lb-ub3} is a decreasing function of $s^{\ast}(\X)$, and hence substituting the above upper bound on $s^{\ast}(\X)$, we have
\[
\peab(\X) \geq 1- \exp\left(-c_1\frac{(mn)^\delta}{\left(c_2+\ln(mn)\right)\ln(mn)} \right),
\]
where $c_1, c_2$ are some positive constants.
Hence $\peab(\X) \tends 1$ as $mn \tends \infty$.
\qed
\subsection{Proof of Theorem \ref{thm:bin_C}}
\label{sect:proof_bin_C}
Recall that $N_{ij}$ is the number of commonly sampled positions in rows $i$ and $j$, given by
\[
N_{ij} = \sum_{k=1}^n 1\left(Y_{ik} \neq e, Y_{jk} \neq e\right).
\]
From the Chernoff bound \cite[Theorem 1]{hoeffding}, we have
\begin{align}
\prob\left(N_{ij}>nr_1(1-\epsilon)^2\right) & \leq \exp\left(-nD\left(r_1(1-\epsilon)^2||(1-\epsilon)^2\right)\right)=\exp(-n\alpha_1), \text{ and } \label{eq:nij1}\\
\prob\left(N_{ij}<nr_2(1-\epsilon)^2\right) & \leq \exp\left(-nD\left(r_2(1-\epsilon)^2||(1-\epsilon)^2\right)\right)=\exp(-n\alpha_2). \label{eq:nij2}
\end{align}
To get a handle on the probability of error, we first analyze it conditioned on the erasure sequence and $\X$.  Let $\bar{E}$ denote the erasure matrix:
\begin{equation*}
\bar{E} = \left[1(Y_{ij}=e)\right]_{m \times n} \in \{0,1\}^{m \times n}.
\end{equation*}

\noindent
{\bf Rows in Same Cluster:}
Consider rows $i,j$ of $\X$ and suppose $I_{ij}=1$, i.e.\ $i,j$ are in the same cluster. We wish to evaluate the probability of error $\prob(d_{ij}\geq d_0\big| I_{ij}=1, \bar{E}, \X)$. In this case,
the random variable $N_{ij}d_{ij}$ is given by 
\[
N_{ij}d_{ij}=\sum_{\substack{\Y_{ik},\Y_{jk}\neq e \\ \X_{ik}=\X_{jk}}}1\left(\Y_{ik}\neq \Y_{jk}\right). 
\]
For any column $k$ such that $\Y_{ik} \neq e, \Y_{jk} \neq e$, the indicator $1(\Y_{ik} \neq \Y_{jk})$ has mean $\mu=2p(1-p)$. Hence,
the above summation has $N_{ij}$ i.i.d \ Bernoulli random variables of mean $\mu$. An application of Chernoff bound \cite[Theorem 1]{hoeffding} yields 
\begin{equation}
\prob\left( d_{ij} \geq d_0 \Big| I_{ij}=1, \bar{E},\X\right) \leq \exp\left(-N_{ij}D(d_0||\mu)\right). \label{eq:prcavg-1-e}
\end{equation}
The bound is independent of $\X$. We only need to take the average of \eqref{eq:prcavg-1-e} with respect to $\bar{E}$. Using \eqref{eq:nij2}, we have
\begin{align}
\prob\left(\hat{I}_{ij}=0\big| I_{ij}=1\right)&=\prob\left( d_{ij} \geq d_0 \Big| I_{ij}=1\right) \nonumber \\
&\leq \exp\left(-n(1-\epsilon)^2r_2D(d_0||\mu)\right) + \exp(-n\alpha_2) \nonumber,\\
&\leq 2 \exp\left(- n \min \left\{ r_2(1-\epsilon)^2 D(d_0 \|| \mu), \alpha_2  \right\} \right) \label{eq:prcavg-1}.
\end{align}

\noindent
{\bf Rows in Different Clusters:}
Next consider the case $I_{ij}=0$, i.e.\ rows $i$ and $j$ are in different clusters. We wish to evaluate $\prob(d_{ij}\leq d_0\big| I_{ij}=0, \bar{E}, \X)$. For $I_{ij}=0$ and fixed $\bar{E}$, $\X$, the random variable $N_{ij}d_{ij}$ is given by 
\begin{equation}
\label{eq:nijdij-0}
N_{ij}d_{ij}=\sum_{\substack{\Y_{ik},\Y_{jk}\neq e \\ \X_{ik}=\X_{jk}}}1\left(\Y_{ik}\neq \Y_{jk}\right) + \sum_{\substack{\Y_{ik},\Y_{jk}\neq e \\ \X_{ik}\neq\X_{jk}}}1\left(\Y_{ik}\neq \Y_{jk}\right) .
\end{equation}
Note that for any column $k$ such that $\Y_{ik} \neq e, \Y_{jk} \neq e$, the indicator $1(\Y_{ik} \neq \Y_{jk})$ has mean 
\begin{itemize}
\item $2p(1-p)=\mu$ if $\X_{ik}=\X_{jk}$, and 
\item $p^2+(1-p)^2=\nu$ if $\X_{ik}\neq \X_{jk}$.
\end{itemize}
Define $S_{ij}$ as the number of columns $k$ such that $\Y_{ik} \neq e, \Y_{jk} \neq e$ and $\X_{ik}\neq \X_{jk}$. Then from \eqref{eq:nijdij-0},we observe that the first sum in \eqref{eq:nijdij-0} has $N_{ij}-S_{ij}$ i.i.d \ Bernoulli random variables of mean $\mu$ and the second sum has $S_{ij}$ i.i.d \ Bernoulli random variables of mean $\nu$, all the random variables being independent. Using the Chernoff bound, we may then write  
\begin{align}
&\prob\left( d_{ij} \leq d_0 \Big| I_{ij}=0,\bar{E}, \X\right) \nonumber \\
& \leq \frac{(1-\nu+\nu e^\theta)^{S_{ij}}(1-\mu+\mu e^\theta)^{N_{ij}-S_{ij}}}{e^{d_0N_{ij}\theta}}, \text{ for }  \theta \leq 0. \label{eq:chern1}
\end{align}
By substituting $h=1-\exp(\theta)$, we can rewrite the above bound as
\begin{align}
\prob\left( d_{ij} \leq d_0 \Big| I_{ij}=0,\bar{E}, \X\right) \leq \frac{(1-\nu h)^{S_{ij}}(1-\mu h)^{N_{ij}-S_{ij}}}{(1-h)^{d_0N_{ij}}}, \text{ for }  0\leq h<1. \label{eq:prcavg-0-x-e}
\end{align}
We are free to choose $0\leq h <1$ in the above bound.  We choose $h$ such that the bound is optimized for the average case $S_{ij}=N_{ij}/2$. For this case, the bound in \eqref{eq:prcavg-0-x-e} reduces to
\[
\left(\frac{(1-\nu h)(1-\mu h)}{(1-h)^{2d_0}}\right)^{N_{ij}/2}.
\]
The value of $h$ that minimizes this bound can be checked to be the smaller root of the quadratic given by \eqref{eq:opt_h}.

Next, we take expectation in \eqref{eq:prcavg-0-x-e} with respect to the erasure sequence $\bar{E}$. Let $s_{ij}$ denote the number of columns $k$ such that $\X_{ik}\neq \X_{jk}$. Then we have, from the Chernoff bound, as in \eqref{eq:nij2},

\begin{equation}
	\prob\left(S_{ij}>s_{ij}r_1(1-\epsilon)^2\right) \leq \exp(-s_{ij}\alpha_1).
\label{eq:Sij}
\end{equation}

Now, from \eqref{eq:prcavg-0-x-e}, we have
\begin{align*}
\prob\left( d_{ij} \leq d_0 \Big| I_{ij}=0,\bar{E}, \X\right) \leq \left(\frac{1-\mu h}{(1-h)^{d_0}}\right)^{N_{ij}}\left(\frac{1-\nu h}{1-\mu h}\right)^{S_{ij}}, \text{ for }  0\leq h<1.
\end{align*}
Now, since $\mu < \nu \leq 1$, we have 
\[
\frac{1-\nu h}{1- \mu h}<1, \text{ for }h\geq 0.
\]
First note that the function $f(h)=(1-\mu h)/(1-h)^{d_0}$ for $h \in [0,1)$ has derivative
\[
f'(h)= \frac{d_0-\mu + \mu h(1-d_0)}{(1-h)^{d_0}}.
\] 
Since $\mu<d_0<1$, $f'(h) >0$ and so $f(h)>f(0)=1$.
Hence $(1- \mu h)/(1-h)^{d_0} > 1$. Now if $S_{ij}>s_{ij}r_2(1-\epsilon)^2$ and $N_{ij}<nr_1(1-\epsilon)^2$, then 
\begin{align*}
\prob\left( d_{ij} \leq d_0 \Big| I_{ij}=0,\bar{E}, \X\right) \leq \left(\frac{1-\mu h}{(1-h)^{d_0}}\right)^{nr_1(1-\epsilon)^2}\left(\frac{1-\nu h}{1-\mu h}\right)^{s_{ij}r_2(1-\epsilon)^2}, \text{ for }  0\leq h<1.
\end{align*}
Combining this with \eqref{eq:Sij} and \eqref{eq:nij1}, we have \\
\begin{align}
\prob\left( d_{ij} \leq d_0 \Big| I_{ij}=0,\X \right) &\leq \left(\frac{1-\mu h}{(1-h)^{d_0}}\right)^{nr_1(1-\epsilon)^2}\left(\frac{1-\nu h}{1-\mu h}\right)^{s_{ij}r_2(1-\epsilon)^2} \nonumber \\
&\quad + \prob\left(S_{ij}<s_{ij}r_2(1-\epsilon)^2\right) + \prob\left(N_{ij}>nr_1(1-\epsilon)^2\right), \nonumber\\
& \leq \left(\frac{1-\mu h}{(1-h)^{d_0}}\right)^{nr_1(1-\epsilon)^2}\left(\frac{1-\nu h}{1-\mu h}\right)^{s_{ij}r_2(1-\epsilon)^2} + \exp(-s_{ij}\alpha_2) + \exp(-n\alpha_1). \label{eq:prcavg-0-x1}
\end{align}
Since $s_{ij}=n_0 X$, where $X$ is Binomial$(t,1/2)$, we have 
\[
\expec\left[\exp(\lambda s_{ij})\right] = \expec \left[\exp(\lambda n_0 X) \right]=\left(\frac{1+\exp(\lambda n_0)}{2}\right)^t.
\] 
Now taking expectation with respect to $\X$ in \eqref{eq:prcavg-0-x1}, we have 
\begin{align}
&P(\hat{I}_{ij}=1\big| I_{ij}=0) =  \prob\left( d_{ij} \leq d_0 \Big| I_{ij}=0 \right) \nonumber \\ 
&\leq \left(\frac{1-\mu h}{(1-h)^{d_0}}\right)^{nr_1(1-\epsilon)^2} \lambda_1(n_0)^t  + \lambda_2(n_0)^t + \exp(-\alpha_1 n) = P_1. \label{eq:prcavg-01}
\end{align}
It remains to show that  
\[
P(\hat{I}_{ij}=1\big| I_{ij}=0) \leq P_2.
\]
This result follows from \eqref{eq:prc_gen} of Theorem \ref{thm:gen_clus} for the general case, which is proved in
Section \ref{sect:proof:thm:gen_clus}.
\qed

\subsection{Proof of Theorem \ref{thm:dmc_knownC}}
\label{sect:proof:thm:dmc_knownC}

For simplicity let $Y_1,...,Y_s$ denote the $s$ samples in block $A_i\times B_j$. Let $\mu_a := q(\cdot | a)$, $a \in \al$ be the transition law of the channel for input $a$ and let $\mu_Y$ denote the empirical probability mass function (PMF) of $Y_1,...,Y_s$. Let $E_{ijs}$ be the error event when the $(i,j)$th block has $s$ samples. For simplicity let ${\cal P}_s$ denote the set of types with denominator $s$ \cite[pp.\ 348]{tmc} and define the set of PMFs:
\[
U_{a,b} := \{\nu: D(\nu\|\mu_b) \leq D(\nu\|\mu_a) \} \cap {\cal P}_s, \quad V_{a,b} := \{\nu: D(\nu\|\mu_b) < D(\nu\|\mu_a) \} \cap {\cal P}_s.
\]

\noindent
{\bf Upper Bound:}
Then
\begin{align*}
\prob(E_{ijs}) & = \frac{1}{|{\al}|} \sum_{a \in \al} \prob(E_{ijs}| a) \\
& \leq \frac{1}{|{\al}|} \sum_{a \in \al} \sum_{b \in {\al}, b \neq a} \prob\left(D(\mu_Y\| \mu_b) \leq D(\mu_Y \| \mu_a) | a\right) \\
&  \leq \frac{1}{|{\al}|} \sum_{a,b \in \al, a \neq b} \sum_{\nu \in U_{a,b}} \exp\left(-s D(\nu \| \mu_a)\right),
\end{align*}
where in the second step we have used the union bound and in the last step we have used \cite[Theorem 11.1.4, pp.\ 354]{tmc}. Let
\begin{equation}
\label{eq:C_1}
C_1 := \lim_{s \tends \infty} -\frac{1}{s}\ln\left(\sum_{a,b \in \al, a \neq b} \sum_{\nu \in U_{a,b}} \exp\left(-s D(\nu \| \mu_a)\right)\right) = \min_{a \neq b} \min_{\{\nu: D(\nu \|\mu_b) \leq D(\nu \| \mu_a)\}} D(\nu \| \mu_a).
\end{equation}
Then for $\delta >0$ small, for $s > s_0(\delta)$, we have
\begin{equation*}
	\prob(E_{ijs}) \leq \frac{\exp\left(-(C_1-\delta) s\right)}{|{\al}|},
\end{equation*}
while for $s \leq s_0$ we can bound this probability by 1. Hence we have from \eqref{eq:eij},
\begin{align}
\prob(E_{ij}^c) = E\left[\prob(E_{ijs}^c)\right] & \geq E\left[1(s\geq s_0)\left(1-\frac{\exp\left(-(C_1-\delta) s\right)}{|{\al}|}\right)\right] \nonumber \\ 
& \quad = E\left[1-\frac{\exp\left(-(C_1-\delta) s\right)}{|{\al}|}\right] - E\left[1(s<s_0)\left(1-\frac{\exp\left(-(C_1-\delta) s\right)}{|{\al}|}\right)\right] \nonumber \\
& \geq E\left[1-\frac{\exp\left(-(C_1-\delta) s\right)}{|{\al}|}\right] - E\left[1(s<s_0)\right] \nonumber \\
& \quad = E\left[1-\frac{\exp\left(-(C_1-\delta) s\right)}{|{\al}|}\right] - \prob(s<s_0). \label{eq:dmc_eijslb}
\end{align}
But for large enough $m_in_j$ using the Chernoff bound \cite[Theorem 1]{hoeffding}, 
\begin{align}
\prob(s < s_0) &\leq \exp\left(-m_in_jD(s_0/m_in_j||1-\epsilon)\right). \label{eq:dmc_Cher}
\end{align}
As $m_in_j\tends \infty$, $D(s_0/m_in_j||1-\epsilon)\tends \ln(1/\epsilon)$. Hence given any $\eta>0$, for large enough $m_in_j$, we have
\[
D( s_0/m_in_j||1-\epsilon) \geq \ln(1/(\epsilon+\eta)). 
\]
Hence, from \eqref{eq:dmc_eijslb} and \eqref{eq:dmc_Cher},
\begin{align}
\prob(E_{ij}^c) &\geq E\left(1-\frac{\exp\left(-(C_1-\delta) s\right)}{|{\al}|}\right) - \left(\epsilon+\eta\right)^{m_in_j} \nonumber \\
&=1-\frac{p_1^{m_in_j}}{|\al|}- \left(\epsilon+\eta\right)^{m_in_j}  \nonumber 
\end{align}
where we have used the fact that $s$ is Binomial$(m_in_j , 1-\epsilon)$ and so the binomial expansion. Note that $\epsilon < p_1$, and hence we can choose $\eta$ so that $\epsilon + \eta < p_1$. Hence we have 
\begin{align}
\prob(E_{ij}^c) \geq 1-\frac{2p_1^{m_in_j}}{|\al|} \nonumber 
\end{align}

Using \eqref{eq:error}, we then have 
\begin{align} 
\peab(\X) &\leq 1- \prod_{i=1,j=1}^{r,t} \left(1-2p_1^{m_i n_j}/|\al|\right) \nonumber \\
&\leq 1- \exp\left(-\frac{4\ln(2)}{|\al|} \sum_{i=1,j=1}^{r,t} p_1^{m_i n_j}\right), \label{eq:dmc_fill_ub} 
\end{align}
where in the last step we have used $1-x \geq \exp(-2\ln(2)x)$ for $x \in [0,1/2]$. Note that for large enough $m_in_j$, we have $p_1^{m_in_j}<1/2$.
But using
\begin{equation*} 
\sum_{i=1,j=1}^{r,t} p_1^{m_i n_j} \leq r t p_1^{s_{\ast}(\X)} \leq \frac{mn}{s_{\ast}(\X)} p_1^{s_{\ast}(\X)},
\end{equation*}
we have,
\begin{align}
\peab(\X) \leq 1-\exp\left(-\frac{4\ln(2)m n p_1^{s_{\ast}(\X)}}{|\al| s_{\ast}(\X)}\right) \label{eq:dmc_ub}.
\end{align}
 The RHS in \eqref{eq:dmc_ub} is a decreasing function of $s_{\ast}(\X)$. Hence if  
\[
s_{\ast}(\X) \geq \frac{\ln(mn)}{\ln(1/p_1)},
\]then 
\[
\peab(\X) \leq 1-\exp\left(-\frac{4\ln(2) \ln(1/p_1)}{|\al|\ln(mn)}\right) \tends 0.
\]

\noindent
{\bf Lower Bound:}
Next we give a lower bound on $\prob(E_{ijs})$. 
If for each $a$ we consider some $b \neq a$, then we get
\begin{align*}
\prob(E_{ijs}) & = \frac{1}{|{\al}|} \sum_{a \in \al} \prob(E_{ijs}| a) \\
& \geq \frac{1}{|{\al}|} \sum_{a \in \al}  \prob\left(D(\mu_Y\| \mu_b) < D(\mu_Y \| \mu_a) | a\right), \\
& \geq \frac{1}{|{\al}|} \sum_{a \in \al}  \sum_{\nu \in V_{a,b}} \frac{\exp\left(-s D(\nu \| \mu_a)\right)}{(s+1)^{|{\al}|}},  \\
& \geq \frac{1}{|{\al}| (m_i n_j +1)^{|{\al}|}} \sum_{a \in \al}  \sum_{\nu \in V_{a,b}} \exp\left(-s D(\nu \| \mu_a)\right), 
\end{align*}
where again we have used \cite[Theorem 11.1.4, pp.\ 354]{tmc} in the third step. Since we are free to choose $b$, we choose it such that
\[
b = \arg \min_{b \neq a} \min_{\{\nu: D(\nu \|\mu_b) < D(\nu \| \mu_a)\}} D(\nu \| \mu_a).
\]
Then we see that
\[
\lim_{s \tends \infty} -\frac{1}{s}\ln\left(\sum_{a \in \al} \sum_{\nu \in V_{a,b}} \exp\left(-s D(\nu \| \mu_a)\right)\right) = C_1.
\]
Hence for $\delta >0$, for $s > s_1(\delta)$, 
\[
\prob(E_{ijs}) \geq \frac{\exp\left(-(C_1+\delta)s\right)}{|{\al}| (m_i n_j +1)^{|{\al}|}}
\]
and for smaller $s$ we use the trivial bound that the probability is non-negative.
Hence we have from \eqref{eq:eij},
\begin{align}
\prob(E_{ij}^c) = E\left(\prob(E_{ijs}^c)\right) & \leq E\left(1-\frac{\exp\left(-(C_1+\delta) s\right)}{|{\al}|(m_i n_j +1)^{|{\al}|}}\right) + \prob(s<s_1) \\
&\leq 1-\frac{p_2^{m_in_j}}{|{\al}|(m_i n_j +1)^{|{\al}|}} + \exp\left(-m_in_jD(s_1/m_in_j||1-\epsilon)\right) \label{eq:dmc_cher1} \\
&\leq 1-\frac{p_2^{m_in_j}}{|{\al}|(m_i n_j +1)^{|{\al}|}} + \epsilon^{m_in_j} \label{eq:dmc_cher2}\\
&\leq 1-\frac{p_2^{m_in_j}}{|{\al}|} + \epsilon^{m_in_j}, \nonumber
\end{align}
where in \eqref{eq:dmc_cher1} we have used the Chernoff Bound \cite[Theorem 1]{hoeffding}, in \eqref{eq:dmc_cher2} we have used the fact that $D(s_1/m_in_j||1-\epsilon) \tends \ln(1/\epsilon)$ monotonically and in the last step we have used $m_in_j \geq 0$. Further, from \eqref{eq:s-ast-x-lb}, we have 
\[
\epsilon^{m_in_j} \leq \frac{p_2^{m_in_j}}{2|\al|},
\]
and hence
\[
\prob(E_{ij}^c) \leq 1-\frac{p_2^{m_in_j}}{2|\al|}.
\]
Using \eqref{eq:error}, we then have 
\begin{align} 
\peab(\X) &\geq 1- \prod_{i=1,j=1}^{r,t} \left(1-\frac{p_2^{m_in_j}}{2|\al|}\right) \nonumber \\
&\geq 1- \exp\left(-\frac{1}{2|\al|}\sum_{i=1,j=1}^{r,t}p_2^{m_in_j}\right) \label{eq:dmc_fill_lb},
\end{align}
where to obtain \eqref{eq:dmc_fill_lb} we have used $1-x \leq \exp(-x)$. Now since 
\begin{equation*} 
\sum_{i=1,j=1}^{r,t} p_2^{m_i n_j} \geq r t p_2^{s^{\ast}(\X)} \geq \frac{mn}{s^{\ast}(\X)} p_2^{s^{\ast}(\X)},
\end{equation*}
we have 
\begin{equation} \nonumber 
\peab(\X) \geq 1- \exp\left(-\frac{mn}{2|\al|s^{\ast}(\X)} p_2^{s^{\ast}(\X)}\right). 
\end{equation}
The RHS above is a decreasing function of $s^{\ast}(\X)$, and hence if
\[
 s^{\ast}(\X) \leq \frac{(1-\zeta)\ln(mn)}{\ln(1/p_2)}, \text{ for some } \delta > 0,
\]
we have
\[
\peab(\X) \geq 1- \exp\left(-\frac{(mn)^\zeta}{2|\al|(1-\zeta)\ln(mn)} \right),
\]
and hence $\peab(\X) \tends 1$ as $mn \tends \infty$.
\qed

\subsection{Proof of Lemma \ref{lem:dineq}}
\label{proof:lem:dineq}
We recall 
\begin{align*}
d_{ub}&=\sum_{p,q}\mu_{pq}/|\al|^2 =\sum_{p,q}\sum_{y\neq z}q(y|p)q(z|q)/|\al|^2. 
\end{align*}
Adding and subtracting the terms corresponding to $y=z$, we have,
\begin{align*}
d_{ub}&=\sum_{p,q}\sum_{y,z}q(y|p)q(z|q)/|\al|^2 \\
&\qquad - \sum_{p,q}\sum_{y}q(y|p)q(y|q)/|\al|^2 \\
&= \left(\sum_{p,y}q(y|p)\right)^2/|\al|^2 - \sum_{p,q}\sum_{y}q(y|p)q(y|q)/|\al|^2.
\end{align*}
Now, $\sum_{p,y}q(y|p)$ is the sum of all entries of the transition probability matrix, and hence is equal to $|\al|$. So we have 
\begin{align}
d_{ub}&=1-\sum_{p,q}\sum_{y}q(y|p)q(y|q)/|\al|^2 \nonumber \\
&=1-\sum_{y}\left(\sum_p q(y|p)\right)^2 \big/|\al|^2. \label{eq:ineq1}
\end{align}
Similarly
\begin{align*}
d_{lb}&=\sum_{p}\mu_{pp}/|\al| =\sum_{p}\sum_{y\neq z}q(y|p)q(z|p)/|\al|. 
\end{align*}
Adding and subtracting the terms coresponding to $y=z$, we have,
\begin{align}
d_{lb}&=\sum_{p}\sum_{y,z}q(y|p)q(z|p)/|\al| - \sum_{p}\sum_{y}q^2(y|p)/|\al| \nonumber \\
&= \sum_{p}\left(\sum_{y}q(y|p)\right)^2/|\al| - \sum_{p}\sum_{y}q^2(y|p)/|\al| \nonumber \\
&=1-\sum_{y}\sum_{p}q^2(y|p)/|\al| \label{eq:ineq2}.
\end{align}
In the last step we have used $\sum_{y}q(y|p)=1$ for the first term.
From \eqref{eq:ineq1} and \eqref{eq:ineq2}, we have
\begin{equation*}
d_{ub}-d_{lb} = \frac{1}{|\al|^2}\sum_{y}\left(|\al|\sum_p q^2(y|p)-\left(\sum_p q(y|p)\right)^2\right).
\end{equation*}
From the Cauchy-Schwarz inequality, 
\[
\left(\sum_p q(y|p)\right)^2 \leq |\al|\sum_p q^2(y|p),
\]
with equality iff $q(y|p)=q(y|q)$ for all $p,q$. The result then follows. \qed

\subsection{Proof of Theorem \ref{thm:gen_clus}}
\label{sect:proof:thm:gen_clus}
We begin with a lemma that provides some useful upper bounds.
\begin{lemma}
\label{lem:1}
Let $Z_1, Z_2, Z_3,\ldots ,Z_t$ be i.i.d\ with mean $\mu$ such that $0\leq Z_i \leq 1,\ \forall \ i\in [1:t]$. Let $m_1, m_2,\ldots, m_t$ and $m$ be positive integers such that $\sum_i m_i = m$. Let 
\[
\beta = \frac{1}{m}\sum_{i=1}^tm_iZ_i.
\]
Then the following hold for sufficiently large $n$.
\begin{enumerate}
\item For $d_0>\mu$, 
\begin{align}
&\prob(\beta > d_0) \nonumber\\ 
& \leq \exp\left(-2(d_0-\mu)^2m^2/(m_1^2+m_2^2+\ldots+m_t^2) \right).  \label{eq:beta1} 
\end{align}
\item For $d_0<\mu$,
\begin{equation}
\label{eq:beta2}
\begin{split}
&\prob(\beta < d_0) \\
& \leq \exp\left(-2(d_0-\mu)^2m^2/(m_1^2+m_2^2+\ldots+m_t^2) \right).
\end{split}
\end{equation}
\item For any positive constant $c$, there exists a positive constant $a$ such that 
\begin{align}
&E\left(\exp(-cm(\beta-d_0)^2)\right) \nonumber \\
& \leq \exp\left(-a(d_0-\mu)^2m^2/(m_1^2+m_2^2+\ldots+m_t^2) \right). \label{eq:beta3}
\end{align}
\end{enumerate}
\end{lemma}

\textit{Proof of Lemma \ref{lem:1}:} \eqref{eq:beta1} and \eqref{eq:beta2} are direct applications of the Chernoff bound \cite[Theorem 2]{hoeffding}. (This particular form is also known as Hoeffding's inequality.) To prove \eqref{eq:beta3}, first assume that $d_0>\mu$. Then
\begin{align}
&E\left(\exp(-cm(\beta-d_0)^2)\right) \nonumber \\
& \leq \prob(\big|\beta-d_0\big|<(d_0-\mu)/2) + \exp(-cm(d_0-\mu)^2/4) \nonumber \\ 
&  \leq \prob(\beta>d_0 - (d_0-\mu)/2) + \exp(-cm(d_0-\mu)^2/4)  \nonumber \\
&  \leq \exp\left(-(d_0-\mu)^2m^2/2(m_1^2+m_2^2+\ldots+m_t^2) \right) \nonumber \\ 
& \quad +\exp(-cm(d_0-\mu)^2/4)\text{ \ \ from \eqref{eq:beta1}}. \nonumber 
\end{align} 
Now, from the Cauchy-Schwarz inequality, we have
\[
m^2/(m_1^2+m_2^2+\ldots +m_t^2) \leq t \leq m.
\]
This gives with $a< \min \{1/2, c/4\}$,
\begin{align}
&E\left(\exp(-cm(\beta-d_0)^2)\right) \nonumber \\
&  \leq \exp\left(-a(d_0-\mu)^2m^2/(m_1^2+m_2^2+\ldots+m_t^2) \right) \label{eq:CS},
\end{align}
for sufficiently large $n$.

To prove \eqref{eq:beta3} in the case $d_0<\mu$, first note that \eqref{eq:beta1} and \eqref{eq:beta2} hold even when the random variables take values in $[-1,0]$. 
Then apply the above result for the random variables $-Z_i,\ i\in [1:t]$. \qed

As in the proof of Theorem \ref{thm:bin_C}, we first analyze the probability of error conditioned on the erasure sequence and $\X$. Let $\bar{E}$ denote the erasure matrix. That is,
\begin{equation*}
\bar{E} = \left(1(Y_{ij}=e)\right)_{m \times n} \in \{0,1\}^{m \times n}.
\end{equation*}

\noindent
{\bf Rows in Same Cluster:}
First consider case when $I_{ij}=1$, i.e.\ $i,j$ are in the same cluster. We wish to evaluate the probability of error $\prob(d_{ij}\geq d_0\big| I_{ij}=1, \bar{E}, \X)$. Define $s_{ij}(p,p)$ as the number of columns $k$ such that $\Y_{ik} \neq e, \Y_{jk} \neq e$ and $\X_{ik}=p$. Clearly,
\[
\sum_{p}s_{ij}(p,p)=N_{ij}.
\]
Note that for such $k$,  the indicator $1(Y_{ik} \neq Y_{jk})$ has mean $\mu_{pp}$. Hence, for $I_{ij}=1$ and a fixed $\bar{E}$, the random variable $N_{ij}d_{ij}$ is given by 
\[
N_{ij}d_{ij}=\sum_{p\in \al}\sum_{\substack{\Y_{ik},\Y_{jk}\neq e \\ \X_{ik}=p=\X_{jk}}}1\left(\Y_{ik}\neq \Y_{jk}\right).
\]
The above summation has $s_{ij}(p,p)$ i.i.d \ Bernoulli random variables of mean $\mu_{pp}$, for each $p \in \al$, all the random variables being independent. Hence the charcterstic function of $N_{ij}d_{ij}$ (for $I_{ij}=1$, fixed $\bar{E}$ and $\X$) is given by
\[
\prod_{p\in \al}(1-\mu_{pp}+\mu_{pp}e^\theta)^{s_{ij}(p,p)}.
\]
Using the Chernoff Bound, we have 
\begin{align}
&\prob\left( d_{ij} \geq d_0 \Big| I_{ij}=1,\bar{E}, \X\right) \nonumber \\
& \leq \frac{\prod_p(1-\mu_{pp}+\mu_{pp}e^\theta)^{s_{ij}(p,p)}}{e^{d_0N_{ij}\theta}}, \text{ for any }\theta\geq 0. \nonumber
\end{align}
By using the inequality $1+x\leq e^x$, we obtain
\begin{align}
&\prob\left( d_{ij} \geq d_0 \Big| I_{ij}=1,\bar{E}, \X\right) \nonumber \\
&  \leq \exp\left(N_{ij}\beta_{ij}(e^\theta-1)-N_{ij}d_0\theta \right),\ \  \theta \geq 0,   \nonumber 
\end{align}
where 
\begin{equation}
	\beta_{ij} = \frac{\sum_{p \in \al}\mu_{pp}s_{ij}(p,p)}{N_{ij}} \label{eq:beta6}.
\end{equation}
Using 
\begin{equation*}
	\theta = \begin{cases}
	&\max\{\ln(d_0/\beta_{ij}), 0\} \text{ if }\beta_{ij}\neq 0\\
	&\infty \text{ if }\beta_{ij}=0,
	\end{cases}
\end{equation*}
we obtain 
\begin{align}
& \prob\left(d_{ij}\geq d_0 \big| I_{ij}=1, \bar{E}, \X\right) \nonumber \\
& \leq \begin{cases}  
\exp \left(N_{ij}({d}_0-\beta_{ij})+N_{ij}{d}_0\ln\left(\frac{\beta_{ij}}{{d}_0}\right)\right) \text{ if }0<\beta_{ij}\leq d_0 \\
1 \text{ if } \beta_{ij}> d_0\\
0\text{ if } \beta_{ij}=0.
\end{cases}  \label{eq:case1} 
\end{align}
For tractability, we further simplify this bound. To do so we note that for
$-1<x\leq 0$ and $0<c<1/2$, the function $f(x)=\ln (1+x)-x + cx^2$ is increasing. This can be seen by noting that 
\[
f'(x) = \frac{x(2c-1+2cx)}{1+x}.
\]
Since $x/(1+x)\leq 0$ and $2c-1+2cx<0$, we have $f'(x) \geq 0$. Hence $\ln (1+x)-x + cx^2\leq 0$ in the interval $-1<x\leq 0$. Now for $0<\beta_{ij}\leq d_0$,  $-1< (\beta_{ij}-d_0)/d_0 \leq 0$, and so 
\begin{equation*}
\ln \left( \frac{\beta_{ij}}{{d}_0}\right) \leq \frac{\beta_{ij}-{d}_0}{{d}_0} - c\left(\frac{\beta_{ij}-{d}_0}{{d}_0}\right)^2. 
\end{equation*}
Using this in \eqref{eq:case1}, for $d_0 \geq \beta_{ij}$, we have  
\begin{equation}
\label{eq:case1-2}
\prob\left(d_{ij}\geq d_0 \big| I_{ij}=1, \bar{E}, \X\right) \leq \exp\left(-cN_{ij}\frac{(\beta_{ij}-d_0)^2}{d_0}\right).
\end{equation}
Taking expectation over $\X$, we obtain
\begin{align}
& \prob\left(d_{ij} \geq d_0 \Big| I_{ij}=1,\bar{E}\right) \nonumber \\
&\leq \prob\left(\beta_{ij} > d_0 \big| I_{ij}=1,\bar{E}\right) \nonumber \\
&\quad  + E\left[\exp\left(-cN_{ij}\frac{(\beta_{ij}-d_0)^2}{d_0}\right)\Big| I_{ij}=1,\bar{E}\right] \label{eq:case1-3} \\
& =: T_1 + T_2. \nonumber
\end{align}
We next bound $T_1$ and $T_2$. 

 For $l \in [1:t]$, let $n_l(\bar{E})$  denote the number of commonly sampled positions for rows $i$ and $j$ in the $l^{th}$ column cluster, i.e.
\[
n_l(\bar{E})=\sum_{k \text{ in cluster }l} 1\left(Y_{ik} \neq e, Y_{jk} \neq e\right).
\]
Note that 
\[
\sum_{l=1}^t n_l(\bar{E}) = N_{ij}
\]
and 
\[
s_{ij}(p,p) = \sum_{l=1}^t n_l(\bar{E})1(\X_{i\{l\}}=p),
\]
where $\X_{i\{l\}}$ is the rating vector of user $i$ in the $l^{th}$ column cluster.
From \eqref{eq:beta6} and the above equation,
\begin{align}
N_{ij}\beta_{ij} &= \sum_{p \in \al}\mu_{pp}\sum_{l=1}^t n_l(\bar{E})1(\X_{i\{l\}}=p) \nonumber \\
&= \sum_{l=1}^tn_l(\bar{E})\sum_{p \in \al}\mu_{pp}1(\X_{i\{l\}}=p) \label{eq:case1-4}, 
\end{align}
where the random variable 
\[
Z_l=\sum_{p \in \al}\mu_{pp}1(\X_{i\{l\}}=p)
\]
takes the value $\mu_{pp}$ with probability $1/|\al|$, for each $p \in \al$. The mean of $Z_l$ is $d_{lb}=\sum_p\mu_{pp}/|\al|$. Further, $Z_l$'s are i.i.d.\ From \eqref{eq:case1-4}, Lemma \ref{lem:1} can be applied to $\beta=\beta_{ij}$, $Z_l$, $m_i=n_i(\bar{E})$. Using \eqref{eq:beta1} of Lemma \ref{lem:1}, we have 
\[
T_1 \leq \exp\left(-a_1(d_0-d_{lb})^2N_{ij}^2/(n_1^2(\bar{E})+n_2^2(\bar{E})+\ldots+n_t^2(\bar{E})) \right)
\]
for some positive constant $a_1$. Similarly using \eqref{eq:beta3} of Lemma \ref{lem:1}, we have
\[
T_2 \leq \exp\left(-a_2(d_0-d_{lb})^2N_{ij}^2/(n_1^2(\bar{E})+n_2^2(\bar{E})+\ldots+n_t^2(\bar{E})) \right)
\]
for some positive constant $a_2$. From \eqref{eq:case1-3}, we then have 
\begin{align*}
\prob\left(d_{ij} \geq d_0 \Big| I_{ij}=1,\bar{E}\right) & \leq T_1 + T_2  \\
& \leq 2 \exp\left(-a(d_0-d_{lb})^2N_{ij}^2/(n_1^2(\bar{E})+n_2^2(\bar{E})+\ldots+n_t^2(\bar{E})) \right)
\end{align*}
for some positive constant $a$ and for sufficiently large $n$.
Using $n_l(\bar{E})\leq n_l$, we can loosen the bound to 
\begin{align*}
& \prob\left(d_{ij} \geq d_0 \Big| I_{ij}=1,\bar{E}\right)  \\
& \leq 2 \exp\left(-a(d_0-d_{lb})^2N_{ij}^2/(n_1^2+n_2^2+\ldots+n_t^2) \right).
\end{align*}
Taking expectation over $\bar{E}$, for $\alpha = r_2(1-\epsilon)^2$ and suitable positive constants $c_1, c_2$ and $c$, we have,
\begin{align}
& \prob\left(d_{ij} \geq d_0 \Big| I_{ij}=1\right) \nonumber  \\
& \leq 2 \exp\left(-a(d_0-d_{lb})^2\alpha^2n^2/(n_1^2+n_2^2+\ldots+n_t^2) \right)\nonumber \\
&\quad + \prob(N_{ij}<n\alpha) \nonumber \\
& \leq 2\exp\left(-c_1n^2/(n_1^2+n_2^2+\ldots+n_t^2) \right) + \exp(-c_2n) \label{eq:case1-5} \\
& \leq c^{\prime} \exp\left(-cn^2/(n_1^2+n_2^2+\ldots+n_t^2) \right) \label{eq:case1-10},
\end{align}
where in \eqref{eq:case1-5} we have used \eqref{eq:nij2}. \eqref{eq:case1-10} is obtained by a similar argument as used to obtain \eqref{eq:CS} using Cauchy Schwarz inequality. 

\noindent
{\bf Rows in Different Clusters:}
Next consider the case $I_{ij}=0$, i.e.\ rows $i$ and $j$ are in different clusters. The bounding technique is similar to the case when $I_{ij}=1$. We wish to evaluate $\prob(d_{ij}\leq d_0\big| I_{ij}=0, \bar{E}, \X)$. Let $s_{ij}(p,q)$ be the number of columns $k$ such that $\Y_{ik} \neq e, \Y_{jk} \neq e, \X_{ik}=p$ and $\X_{jk}=q$. Then for a $I_{ij}=0$ and fixed $\bar{E}$, $\X$, the random variable $N_{ij}d_{ij}$ is given by 
\[
N_{ij}d_{ij}=\sum_{p,q\in \al}\sum_{\substack{\Y_{ik},\Y_{jk}\neq e \\ \X_{ik}=p,\X_{jk}=q}}1\left(\Y_{ik}\neq \Y_{jk}\right).
\]
The above summation has $s_{ij}(p,q)$ i.i.d \ Bernoulli random variables of mean $\mu_{pq}$, for each $(p, q) \in \al^2$, all the random variables being independent. 
Using the Chernoff Bound, we may then write  
\begin{align}
&\prob\left( d_{ij} \leq d_0 \Big| I_{ij}=0,\bar{E}, \X\right) \nonumber \\
&  \leq \frac{\prod_{p,q}(1-\mu_{pq}+\mu_{pq}e^\theta)^{s_{ij}(p,q)}}{e^{d_0N_{ij}\theta}} \theta \leq 0. \nonumber
\end{align}
By using the inequality $1+x\leq e^x$ we obtain
\begin{align}
&\prob\left( d_{ij} \leq d_0 \Big| I_{ij}=0,\bar{E}, \X\right) \nonumber \\
& \leq \exp\left(N_{ij}\beta_{ij}(e^\theta-1)-N_{ij}d_0\theta \right),\text{ for any }\theta\leq 0,   \nonumber 
\end{align}
where 
\begin{equation}
	\beta_{ij} = \frac{\sum_{p,q \in \al}\mu_{pq}s_{ij}(p,q)}{N_{ij}}. \label{eq:beta7}
\end{equation}
Using $\theta=\min\{\ln(d_0/\beta_{ij}), 0\}$, we obtain 
\begin{align}
& \prob\left(d_{ij}\leq d_0 \big| I_{ij}=0, \bar{E}, \X\right) \nonumber \\
& \leq \begin{cases}  
\exp \left(N_{ij}({d}_0-\beta_{ij})+N_{ij}{d}_0\ln\left(\frac{\beta_{ij}}{{d}_0}\right)\right) \text{ if }\beta_{ij}\geq d_0 \\
1 \text{ if } \beta_{ij}< d_0.
\end{cases}  \label{eq:case2} 
\end{align}

But $s_{ij}(p,q) \leq N_{ij}$, and so 
\[
\frac{\beta_{ij}}{d_0} \leq \frac {\sum_{p,q}\mu_{pq}}{d_0} = 1+s,
\]
where $s$ is defined as
\[
s := \frac{\sum_{p,q}\mu_{pq}}{d_0} - 1.
\]
So for $\beta_{ij}\geq d_0$, we have $0\leq (\beta_{ij}-d_0)/d_0\leq s$. 
But for any $0<c<1/(2(1+s))$, the function $f(x)=\ln (1+x) -x +cx^2$ is a decreasing function on $[0,s]$. So we have the following  
\begin{equation*}
\ln \left( \frac{\beta_{ij}}{{d}_0}\right) \leq \frac{\beta_{ij}-{d}_0}{{d}_0} - c\left(\frac{\beta_{ij}-{d}_0}{{d}_0}\right)^2. 
\end{equation*}
Using this in \eqref{eq:case2}, for $\beta_{ij} \geq d_0$, we have  
\begin{equation}
\label{eq:case2-2}
\prob\left(d_{ij}\leq d_0 \big| I_{ij}=0, \bar{E}, \X\right) \leq \exp\left(-cN_{ij}\frac{(\beta_{ij}-d_0)^2}{d_0}\right).
\end{equation}
Taking expectation over $\X$, we obtain
\begin{align}
& \prob\left(d_{ij} \leq d_0 \Big| I_{ij}=0,\bar{E}\right) \nonumber \\
& \leq \prob\left(\beta_{ij} < d_0 \Big|I_{ij}=0,\bar{E}\right) \nonumber \\
& \quad + E\left[\exp\left(-cN_{ij}\frac{(\beta_{ij}-d_0)^2}{d_0}\right)\Big|I_{ij}=0,\bar{E}\right]. \label{eq:case2-3}
\end{align}
Then we follow the same line of arguments as in the case when $I_{ij}=1$. Note that now 
\begin{align}
N_{ij}\beta_{ij} &= \sum_{p \in \al}\mu_{pp}\sum_{l=1}^t n_l(\bar{E})1(\X_{i\{l\}}=p) \nonumber \\
&= \sum_{l=1}^tn_l(\bar{E})\sum_{p \in \al}\mu_{pp}1(\X_{i\{l\}}=p) \label{eq:case2-4}, 
\end{align}
where the random variable 
\[
Z_l=\sum_{p,q \in \al}\mu_{pq}1(\X_{i\{l\}}=p)1(\X_{j\{l\}}=q)
\]
takes the value $\mu_{pq}$ with probability $1/|\al|^2$. The mean of $Z_l$ is $d_{ub}=\sum_{p,q}\mu_{pq}/|\al|^2$. Further, $Z_l$'s are i.i.d\ Applying Lemma \ref{lem:1} and \eqref{eq:nij2} as in the case of $I_{ij}=1$, we  again have 
\begin{align*}
& \prob\left(d_{ij} \leq d_0 \Big| I_{ij}=0\right) \nonumber  \\
& \leq c^{\prime} \exp\left(-cn^2/(n_1^2+n_2^2+\ldots+n_t^2) \right).
\end{align*}
Since there are at most $m(m-1)/2$ pairs of rows, the result follows by the union bound. \qed


\begin{thebibliography}{10}

\bibitem{netflix}
http://www.netflixprize.com/

\bibitem{us_isit09}
S.T. Aditya, O. Dabeer, B.K. Dey, ``A Channel Coding Perspective of Recommendation Systems," \emph{ISIT,} 2009, Seoul, Korea.

\bibitem{surveypaper}  	
G.\ Adomavicius, A.\ Tuzhilin, ``Toward the Next Generation of Recommender Systems: A Survey of the State-of-the-Art and Possible Extensions,'' \emph{IEEE Tran.\ Knowledge and Data Engineering}, vol.\ 17, no.\ 6, pp.~734-749, June 2005.


\bibitem{candes_plan}
E.\ J.\ Candes, Y.\ Plan, ``Matrix completion with noise," {\it Arxiv Preprint arXiv:0903.3131}, 2009

\bibitem{recht2}
E.\ J.\ Candes, B.\ Recht, ``Exact Matrix Completion via Convex Optimization,'' {\it Arxiv Preprint arXiv:0805.4471}, 2008

\bibitem{candes_tao}
E.\ J.\ Candes, T.\ Tao, ``The Power of Convex Relaxation: Near-Optimal Matrix Completion," {\it Arxiv Preprint arXiv:0903.1476}, 2009

\bibitem{chakrabarti}
S.\ Chakrabarti, ``Mining the Web," Morgan Kaufmann Publishers, San Fransisco, 2003

\bibitem{tmc}
T.M. Cover, J.A. Thomas, \emph{Elements of Information Theory,} Second Edition, John Wiley \& Sons, 2006

\bibitem{sp.issue}
A.\ Felfernig, G.\ Friedrich, L.\ Schmidt-Thieme, ``Guest Editor's Introduction: Recommender Systems,'' \emph{IEEE Intelligent Systems}, vol.\ 22 no.\ 3, pp.~18-21, May 2007   

\bibitem{hoeffding}
W. Hoeffding, ``Probability Inequalities for Sums of Bounded Random Variables," \emph{Journal of the American Statistical Association}, vol.\ 58, no.\ 301, pp.~ 13-30, Mar 1963.


\bibitem{montanari}
R. Keshavan, A. Montanari, S. Oh, ``Learning low rank matrices from O(n) entries,'' \emph{Allerton} 2008.

\bibitem{keshavan_isit09}
R. Keshavan, A. Montanari, S. Oh, ``Matrix Completion from a Few Entries,'' \emph{ISIT} 2009, Seoul, Korea.

\bibitem{koren} Y.\ Koren, ``Factorization Meets the Neighborhood: a Multifaceted Collaborative Filtering Model," {\em ACM Int. Conference on Knowledge Discovery and Data Mining (KDD'08)}, 2008

\bibitem{bresler_admira}
K.\ Lee and Y.\ Bresler, ``ADMiRA: Atomic Decomposition for Minimum Rank Approximation," {\it Arxiv Preprint  arXiv:0905.0044}, 2009

\bibitem{mitzenmacher}
M. Mitzenmacher, E. Upfal, \emph{Probability and Computing: Randomized Algorithms and Probabilistic Analysis,} Cambridge University Press, 2005

\bibitem{recht1}
B. Recht, M. Fazel, P. A. Parrilo, ``Guaranteed minimum rank
solutions to linear matrix equations via nuclear norm minimization,'' submitted to \emph{SIAM Review}, 2007


   
\end{thebibliography}
\end{document}